\documentclass[aps,prl,twocolumn,superscriptaddress,footinbib]{revtex4-1}  
\usepackage{graphicx}  
\usepackage{dcolumn}   
\usepackage{bm}        
\usepackage{amssymb}   
\usepackage{amsmath}   
\usepackage{extarrows}

\newcommand{\hide}[1]{}
\renewcommand{\ao}{\hat{a}}
\renewcommand{\aa}{\hat{a}^\dag}

\newcommand{\co}{\hat{c}}
\newcommand{\ca}{\hat{c}^\dag}
\newcommand{\no}{\hat{n}}
\newcommand{\ra}{\rangle}
\newcommand{\la}{\langle}
\newcommand{\bn}{{\bm n}}
\newcommand{\rd}{{\mathrm d}}
\newcommand{\be}{\begin{equation}}
\newcommand{\ee}{\end{equation}}
\newcommand{\bes}{\begin{eqnarray}}
\newcommand{\ees}{\end{eqnarray}}
\newcommand{\twopihbar}{}
\begin{document}

\title{High-temperature nonequilibrium Bose condensation induced by a hot needle}
\author{Alexander~Schnell} 
\email[Electronic address: ]{schnell@pks.mpg.de}
\affiliation{Max-Planck-Institut f{\"u}r Physik komplexer Systeme, N{\"o}thnitzer Stra\ss e 38, 01187 Dresden, Germany}
\author{Daniel~Vorberg} 
\affiliation{Max-Planck-Institut f{\"u}r Physik komplexer Systeme, N{\"o}thnitzer Stra\ss e 38, 01187 Dresden, Germany}
\author{Roland~Ketzmerick} 
\affiliation{Max-Planck-Institut f{\"u}r Physik komplexer Systeme, N{\"o}thnitzer Stra\ss e 38, 01187 Dresden, Germany}
\affiliation{Technische Universit{\"a}t Dresden, Institut f{\"u}r Theoretische Physik and Center for Dynamics, 01062 Dresden, Germany}
\author{Andr{\'e}~Eckardt} 
\email[Electronic address: ]{eckardt@pks.mpg.de}
\affiliation{Max-Planck-Institut f{\"u}r Physik komplexer Systeme, N{\"o}thnitzer Stra\ss e 38, 01187 Dresden, Germany}

\date{\today}

\begin{abstract}
We investigate theoretically a one-dimensional ideal Bose gas that is driven into
a steady state far from equilibrium via the coupling to two heat baths: a global 
bath of temperature $T$ and a ``hot needle'', a bath of temperature $T_h\gg T$ 
with localized coupling to the system. Remarkably, this system features a crossover to
finite-size Bose condensation at temperatures $T$ that are orders of magnitude 
larger than the equilibrium condensation temperature. This counterintuitive effect
is explained by a suppression of long-wavelength excitations resulting from the 
competition between both baths. Moreover, for sufficiently large needle 
temperatures ground-state condensation is superseded by condensation into an 
excited state, which is favored by its weaker coupling to the hot needle. Our 
results suggest a general strategy for the preparation of quantum degenerate 
nonequilibrium steady states with unconventional properties and at large 
temperatures. 
\end{abstract}

\maketitle

\emph{Introduction.}---%
In thermal equilibrium the state of a system is strongly restricted by the laws
of thermodynamics. Irrespective of the details of the environment, it is 
characterized by a few state variables only, like temperature or chemical 
potential. This is not the case anymore, when the system is driven far away from 
equilibrium and the exciting question arises, whether the freedom to prepare 
nonthermal states of matter can be used to manipulate the properties of many-body 
quantum systems in a controlled fashion. Recently,
this led to various interesting directions of research. In transient states,
dynamically induced Bose condensation  \cite{RigolMuramatsu04,VidmarEtAl15},
light-induced superconductivity \cite{MitranoEtAl16}, and dynamical phase 
transitions \cite{HeylPolkovnikovKehrein13, JurcevicEtAl16, FlaeschnerEtAl16} 
were studied.
Nonequilibrium steady states, whose properties depend on the initial condition,
can occur in many-body localized isolated systems
\cite{BaskoAleinerAltshuler05, NandkishoreHuse15,SchreiberEtAl15, SmithEtAl16,
ChoiEtAl16}, including Floquet (i.e.~time-periodically driven) systems
\cite{PonteEtAl15, LazaridesEtAl15} such as discrete time crystals
\cite{KhemaniEtAl16, vonKeyserlingkEtAl16, ElseBauerNayak16, ZhangEtAl17,
ChoiEtAl17}. Floquet engineering, the coherent control of isolated systems by
periodic driving on long (but finite) time scales, was very successfully used in
atomic quantum gases, e.g. for the realization of artificial magnetic fields
\cite{BukovEtAl15, Eckardt17}. Also nonequilibrium steady states of
driven-dissipative many-body systems attracted considerable attention, including 
open Floquet systems \cite{TsujiEtAl09, VorbergEtAl13, VorbergEtAl15,
FoaTorresEtAl14, SeetharamEtAl15, DehghaniEtAl15, GoldsteinEtAl15, IadecolaEtAl15, 
ShiraiEtAl16, QinHofstetter17} and photonic systems \cite{CarusottoCiuti13}, where 
i.a.\ the question was studied in how far Bose condensation can be distinguished
from lasing \cite{KlaersEtAl10, ByrnesKimYamamoto14, LeymannEtAl17}.

\begin{figure}[h!]
	\includegraphics{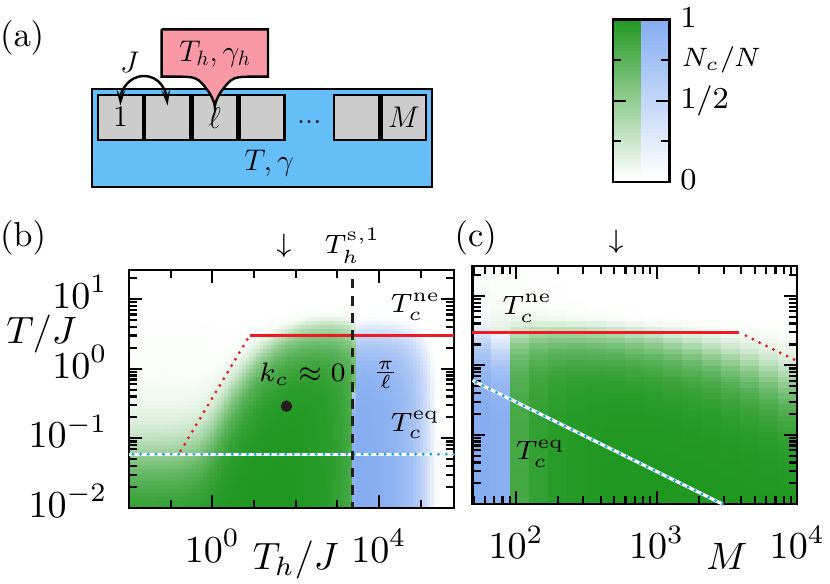}
\caption{(color online)
(a) Tight-binding chain with $N=nM$ bosons on $M$ sites and tunneling 
parameter $J$, coupled with strengths $\gamma$ and $\gamma_h$ to a global bath
of temperature $T$ and to a ``hot needle'' of temperature $T_h$ at site $\ell$,
respectively. 
(b,c) Condensate fraction $N_c/N$ indicated by green (blue) shading for
ground-state (excited-state) condensation in the mode $k_c\approx 0$
($k_c\approx \pi/\ell$) versus $T$ and $T_h$ or $M$; for $\ell=5$,
$\gamma_h=0.5\gamma$, $n=3$ and (b) $M=500$ or (c) $T_h=60J$. 
Estimated temperature $T_c^\text{ne}$ for 1D-like (3D-like) Bose condensation 
plotted as red dotted (solid) lines. Blue-white dotted lines give equilibrium 
condensation temperature $T_c^\text{eq}$ (for $\gamma_h=0$). Black dashed line 
gives estimated needle temperature $T_h^\text{s,1}$, where excited-state 
condensation sets in. Black arrows indicate the parameters where (b) and (c) coincide.
}
\label{fig:mainfig}
\end{figure}

In this work we investigate the nonequilibrium steady state of a quantum gas in
contact with two heat baths of different temperature. In particular, we consider
a one-dimensional (1D) ideal Bose gas that is coupled globally to an environment
of temperature $T$ and driven into a steady state far from equilibrium via the 
additional coupling to a ``hot needle'', a local bath of temperature
$T_h\gg T$ [Fig.~\ref{fig:mainfig}(a)]. We find the surprising effect that a crossover to
Bose condensation can occur when both temperatures $T$ and $T_h$
are orders of magnitude larger than the temperature where (finite-size) condensation occurs in 
equilibrium. We explain this behavior by a suppression of long-wavelength excitations
resulting from the competition between both baths. Moreover, we observe that
for sufficiently large needle temperatures Bose condensation occurs in an excited 
state of the system, which provides a better decoupling from the hot needle. This 
intriguing phenomenon bears resemblance to the quantum Zeno effect.

\emph{System and model.}---%
Let us consider a 1D system of $N$ noninteracting bosons 
that tunnel between adjacent sites of a tight-binding chain of length $M$
[Fig.~\ref{fig:mainfig}(a)]. The Hamiltonian reads
\be
\hat{H} = -J \sum_{i=1}^{M-1} \! \big( \aa_{i+1}\ao_i + \text{h.c.}\big)
      = \sum_k\varepsilon_k \no_k.
\ee
Here $J$ is the tunneling parameter and $\ao_i$ the bosonic annihilation 
operator at lattice site $i$. The dimensionless wave numbers $k=\pi\nu/(M+1)$ 
with $\nu=1,\ldots,M$ characterize the single-particle energy 
eigenmodes with energy $\varepsilon_k=-2J\cos(k)$, wave function
$\langle i|k\rangle=\sqrt{2/(M+1)}\sin(k i)$ (describing a superposition of 
states with quasimomenta $k$ and $-k$), and number operator $\no_k=\ca_k\co_k$ 
with $\co_k=\sum_i\langle k|i\rangle \ao_i$. The eigenstates of the 
Hamiltonian are Fock states $|\bn\ra$ labeled by the vector $\bn$ of 
occupation numbers $n_k$.

A heat bath $b$ is modeled as a collection of harmonic oscillators in thermal 
equilibrium with temperature $T_b$ that couple to a single-particle system 
operator $\hat{v}^{(b)}\equiv\sum_{qk}v_{qk}^{(b)}\ca_q\co_k$. 
In the limit of weak system-bath coupling (small compared to $ \min_{k\neq q}\lbrace| \Delta_{qk} |\rbrace \approx 1.5J/M^2$, with 
$\Delta_{qk}\equiv\varepsilon_q-\varepsilon_k$), the bath induces quantum jumps between 
the energy eigenstates $|\bn\ra$ of the system, where a boson is transferred from 
mode $k$ to mode $q$ with rate $(n_q+1)n_k R^{(b)}_{qk}$. Here the dependence
on the occupation $n_q$ reflects the bosonic quantum statistics.
The single-particle rate $R^{(b)}_{qk}$ is obtained within the rotating-wave
Born-Markov approximation and is given by the golden-rule-type expression
$R^{(b)}_{qk}=\frac{2\pi}{\hbar}|v^{(b)}_{qk}|^2 J_b(\Delta_{qk}) 
[\exp{(\Delta_{qk}/k_\textup{B}T_b)}-1]^{-1}$ \cite{BreuerPetruccione}. We choose ohmic baths with spectral density
$J_b(\Delta)\propto \Delta$. 
Setting
$\hbar=k_\textup{B}=1$ from now on, 
 the rates take the form
\be\label{eq:rates}
R^{(b)}_{qk}= \twopihbar f^{(b)}_{qk} \gamma_b^2
                \frac{\Delta_{qk}}{e^{\Delta_{qk}/T_b}-1} 
        \xlongrightarrow{T_b \gg |\Delta_{qk}| }
        \twopihbar f^{(b)}_{qk} \gamma_b^2T_b,
\ee
with the dimensionless coupling strength $\gamma_b$ and factor
$f^{(b)}_{qk}\propto|v^{(b)}_{qk}|^2$.
When deriving analytical estimates, the asymptotic expression for $T_b\gg|\Delta_{qk}|$ will be employed for the 
hot bath and for long-wavelength modes $k,q\ll 1$, otherwise the full expression is used.
We also define the rate asymmetry
\be\label{eq:A}
A^{(b)}_{qk} = R^{(b)}_{qk}-R^{(b)}_{kq} 
    = - \twopihbar f^{(b)}_{qk} \gamma_b^2 \Delta_{qk}.
\ee
We will consider two baths, $R_{qk} = R_{qk}^{(g)} + R_{qk}^{(h)}$, 
a global bath $g$ of temperature $T$ and coupling strength $\gamma$ as well as 
a hot local bath $h$ at site $\ell$ (the hot needle) of temperature $T_h$ and 
coupling strength $\gamma_h$ [Fig.~\ref{fig:mainfig}(a)]. 
The hot needle couples to the operator $\hat{v}^{(h)}=\aa_\ell\ao_\ell$ so that
$f^{(h)}_{qk}=4\sin^2(\ell q)\sin^2(\ell k)$, whereas the global bath is 
modeled by a collection of local baths of temperature $T_g=T$, each coupling 
to the occupation $\aa_i\ao_i$ of one site with strength $\gamma/\sqrt{M}$, 
so that $R^{(g)}_{qk}=\sum_i R^{(gi)}_{qk}$ gives
$f^{(g)}_{qk}=\sum_i4\sin^2(i q)\sin^2(i k)/M\simeq 1$.

In order to treat large systems 
and as a starting point for analytical approximations, we employ the meanfield approximation
$\la\no_q\no_k\ra\approx\la\no_q\ra\la\no_k\ra$. It gives rise to a closed set 
of nonlinear kinetic equations for the mean occupations $\la\no_k\ra$ from 
which we obtain the steady state, 
\be
\frac{\rd \la\no_k\ra}{\rd t} = \sum_q\! \big[ A_{kq}\la\no_q\ra\la\no_k\ra 
                + R_{kq}\la\no_q\ra-R_{qk}\la\no_k\ra\big] = 0.
    \label{eq:mf}
\ee
In the supplemental material we compare meanfield with exact Monte-Carlo results 
for $M=50$ and find excellent agreement \cite{Supp} (see also
Ref.~\cite{VorbergEtAl15} for a detailed description of both methods). Note that 
for a fixed ratio $\gamma_h/\gamma$, the steady state does not depend on $\gamma$.
After having defined the system, we are now in the position to compute the 
steady-state mean occupations $\la\no_k\ra$ from Eq.~(\ref{eq:mf}). 

\emph{Finite-size equilibrium condensation.}---%
Let us 
first recapitulate the equilibrium case, where the system is coupled to a 
single bath of temperature $T$ only. Here one recovers the familiar
grand-canonical mean occupations
 $\la\no_k\ra_\text{eq}=[e^{(\varepsilon_k-\mu)/T}-1]^{-1}$, where the 
chemical potential $\mu$ has to be adjusted so that 
$\sum_q\la\no_k\ra_\text{eq}=N$. In the thermodynamic limit, $M\to\infty$ at 
constant density $n=N/M$, thermal fluctuations prevent the formation of a 
Bose condensate in a 1D system at finite temperature. However, 
for a finite system size $M$, a crossover into a Bose condensed regime with a 
relative occupation $\la\no_{k_c}\ra/N$ of order one in 
the ground state $k_c=\pi/(M+1)$ occurs when $T$ reaches the condensation 
temperature $T_c^\text{eq}\approx 8.3\,nJ/M$, which we define as the 
temperature at which half of the particles occupy the single-particle ground 
states \cite{Supp}.  

\emph{High-temperature nonequilibrium condensation.}---%
Turning to the nonequilibrium situation with both the global bath and 
the hot needle present, we have to compute the steady state by solving
Eq.~(\ref{eq:mf}) numerically. Figure~\ref{fig:mainfig}(b) shows the 
condensate fraction $N_c/N$, with occupation $N_c$ of the most populated mode
$k_c$, versus both temperatures $T$ and $T_h$ 
for a system of 500 sites with $n=3$, $\ell=5$, and $\gamma_h/\gamma=0.5$. 
For small needle temperatures, $T_h\lesssim T$, we find a crossover into a
Bose-condensed regime, roughly when the global temperature $T$ falls below the 
equilibrium value $T_c^\text{eq}$ (blue-white dotted line). However, when the
needle temperature is \emph{increased} further, an astonishing effect occurs: 
The global temperature at which condensation occurs \emph{increases} by almost
two orders of magnitude until it reaches a saturation value. Thus, for an 
environment well above the equilibrium condensation temperature $T_c^{\text{eq}}$, 
coupling the system to a second, even hotter local bath (the hot needle) can
induce Bose condensation. When the needle temperature is increased even further, 
we can observe another intriguing effect: the condensate is suddenly 
formed in an excited state, $k_c\approx\pi/\ell$, as indicated by the 
color code. Only for very large needle temperatures, condensation eventually 
breaks down completely. 

In Fig.~\ref{fig:mainfig}(c) the condensate fraction is plotted versus global 
temperature $T$ and system size $M$. One sees that up to 
large system sizes of about $10^3$ sites, condensation occurs at a large 
condensation temperature $T_c^\text{ne}$ that is practically independent of the 
system size. This behavior is reminiscent of the physics of Bose condensation 
in a three-dimensional (3D) system. Only at even larger system sizes, the 
condensation temperature decreases with $M$ resembling the equilibrium behavior 
in 1D. In the limit of small $M$ ($\lesssim 100$), again excited-state 
condensation in the mode $k_c\approx\pi/\ell$ is found. 

\begin{figure}[t]
    \includegraphics{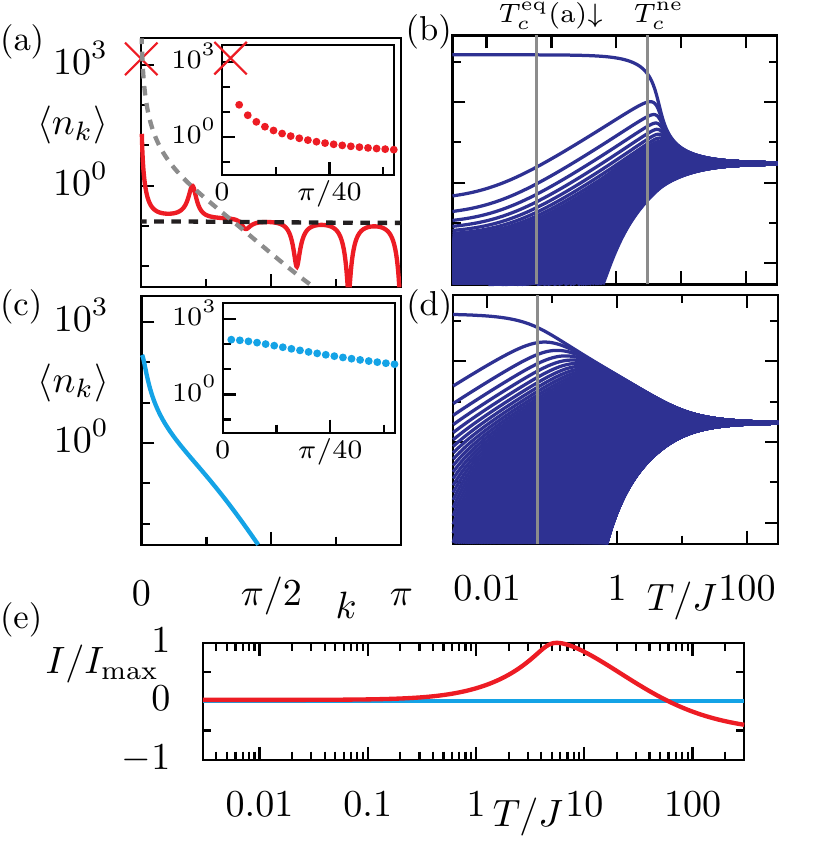}
    \caption{(color online) (a,b) Mean occupations $\langle n_k\rangle$ for 
    the parameters of Fig.~\ref{fig:mainfig} with $T_h=60J$ and $M=500$, (a) versus
    $k$ [for $T=0.29J$, black dot in Fig.~\ref{fig:mainfig}(b)] and (b) versus
    $T$. In (a) the crosses indicate the condensate 
    occupation, the gray (black) dashed line in the main panel shows thermal 
    distributions for the temperature $T$ ($T_h$); the inset shows $\la\no_k\ra$
    for small $k$.
    (c,d) Like (a,b), but for the equilibrium situation with $\gamma_h/\gamma=0$.
    Occupations obey the Bose-Einstein distribution.
    (e) Heat current $I$ from the needle through the system into the global 
    bath versus $T$ for the nonequilibrium steady state [red line, parameters    
    like in (b)], the blue line shows the trivial equilibrium value 
    $I=0$.}
	\label{fig:occups}
\end{figure}

\emph{Momentum distribution.}---%
In order to obtain a better understanding of the intriguing behavior observed 
in Figs.~\ref{fig:mainfig}(b) and (c), let us have a look at 
the full momentum distribution $\la\no_k\ra$. It is plotted in
Fig.~\ref{fig:occups}(a) for the parameters indicated by the black dot in
Fig.~\ref{fig:mainfig}(b).
The occupation of the condensate formed in the ground state is indicated by a 
red cross and the occupation of all other modes by a red line. We find an 
unconventional nonmonotonous behavior of $\la\no_k\ra$ with equidistant peaks or 
dips. They are located around those wave numbers $\kappa_\alpha = \pi\alpha/\ell$ 
with $\alpha=0,1,\ldots,\ell$ that decouple from the hot needle,
$f^{(h)}_{q\kappa_\alpha}=f^{(h)}_{\kappa_\alpha q}=0$. For momenta
$k\approx\kappa_\alpha$ the distribution $\la\no_{k}\ra$ approximately follows a 
thermal distribution with temperature $T$ (dashed gray line). 
Between the momenta $\kappa_\alpha$, the distribution roughly follows the thermal 
distribution associated with the hot temperature $T_h$ of the needle (dashed 
black line), which is rather flat. This behavior can be explained by noting that for
$\vert\Delta_{qk}\vert \ll T_b$ the rates (\ref{eq:rates}) become proportional to 
the bath temperature, so that the occupations $\la \hat{n}_k\ra$ are dominated by 
the hot bath with $T_h\gg T$, except for momenta near $\kappa_\alpha$ that almost 
decouple from the needle.

This discussion gives us already an idea of the mechanism behind the
high-temperature condensation induced by the needle. Namely, the width of the 
peak of $\la\no_k\ra$ at $k=0$ is now determined by the competition between the 
global bath and the hot needle. The estimate 
$w=(\gamma/\ell\gamma_h)\sqrt{T/T_h}$ for the peak width can be obtained from 
requiring that $R^{(h)}_{qk} / R^{(g)}_{qk} \approx {\gamma_h^2}T_h f^{(h)}_{qk}/(\gamma^2 T)  \lesssim 1$ for $k<w$ and all $q$. It can be small 
compared to the width of the thermal distribution at temperature
$T$ [Fig.~\ref{fig:occups}(c)]. In this way long-wavelength excitations, which 
destroy Bose condensation in 1D equilibrium systems for temperatures above
$T_c^\text{eq}$, are reduced. The effect of Bose condensation and the 
dramatic increase of the condensation temperature $T_c$ induced by the 
hot needle can clearly be observed in Figs.~\ref{fig:occups}(b) and (d), 
showing the $T$ dependence of the occupations $\la\no_k\ra$ for a system with 
and without coupling to the needle, respectively. 

\emph{Estimating the condensation temperature.}---%
Based on our qualitative discussion, we can estimate the nonequilibrium
condensation temperature $T_c^\text{ne}$. In the condensate regime, where a 
large fraction $N_c/N$ of the particles occupy the ground-state mode
$k_0=\pi/(M+1)$, the occupation of excited modes $k\ll 1$ in the vicinity of
$k_0$ is approximately given by
$\la\no_k\ra \approx  N_c w^2 /(2M k^2)$ \cite{Supp}.
We obtain this expression from Eq.~(\ref{eq:mf}) as follows: We assume that all 
particles occupy long-wavelengths modes $k\ll1$ (which is a valid approximation 
for $T\ll T_c^{\text{ne}2}$, see next paragraph) and neglect the coupling of the 
hot bath to the condensate mode $k_0$, but not to other long-wavelength modes.
The condensation temperature, defined by $N_c=N/2$, can then be estimated 
to read
$T_c^\text{ne1}\approx 30.6 \frac{\ell^2\gamma_h^2T_h}{\gamma^2} \,\frac{1}{M}$.
It is plotted as red dotted line in Figs.~\ref{fig:mainfig}(b) and (c) and
agrees well with the observed behavior. Like in equilibrium in 1D, also the 
nonequilibrium condensation relies on the infrared cutoff given by the inverse 
system size.

However, in contrast to equilibrium, in the nonequilibrium steady state Bose 
condensation can also be destroyed by increasing the occupation of modes with 
large momenta $q$, which couple to the condensate via the rates induced by the 
global bath. This effect can be estimated by considering the regime
$T\ll T_c^\text{ne1}$, where the total occupation of excited long-wavelength 
modes is suppressed by the finite system size $M$, so that we can approximate the 
occupation of excited states by the flat distribution
$\la\no_q\ra\approx (N-N_c)/M$ induced by the dominant hot-needle bath [dashed black line in Fig.~\ref{fig:occups}(a)].
Considering the coupling of the so-occupied excited states to the condensate via 
the global bath, we find $N-N_c=TM/(2J)$  \cite{Supp}.
From requiring $N_c=N/2$, one finds the condensation temperature
$T_c^\text{ne2} \approx nJ$, which is plotted as red solid line in
Fig.~\ref{fig:mainfig}(b) and (c). The fact that it does not depend on the 
system size, reflects similarity to the break-down of Bose condensation in a 3D 
system in equilibrium, which is not driven by long-wavelength modes either. 

\begin{figure}
	\includegraphics{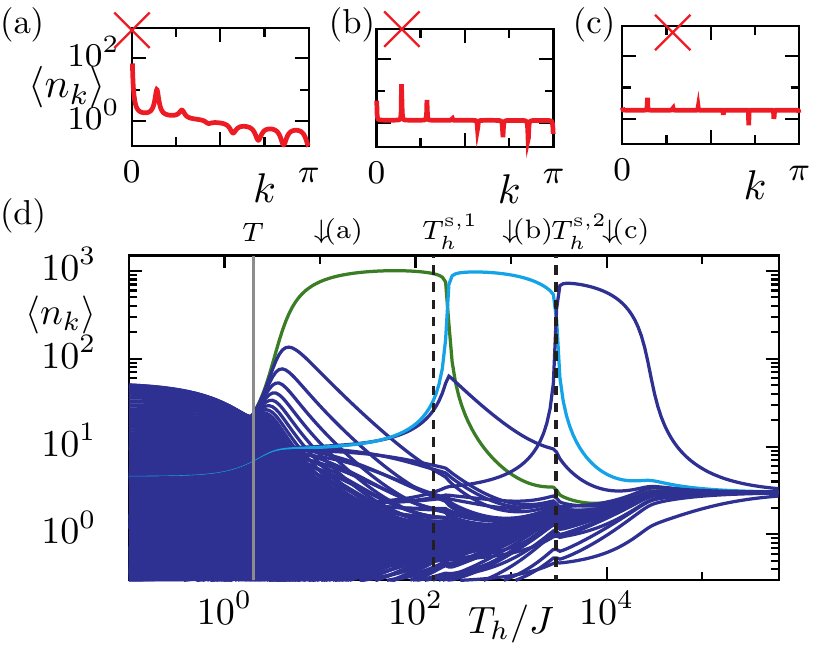}
	\caption{(color online) Mode occupations (d) versus needle temperature $T_h$ for $T=2J$,
$\ell=7$, $M=500$, $n=3$, and $\gamma_h/\gamma = 1$. With increasing $T_h$ the system
passes from a state without Bose condensate through a sequence of states with a
condensate in the states $k_0\approx 0$ (green line), $k_1\approx\pi/\ell$ 
(light blue line), and $k_2\approx 2\pi/\ell$ (violet line), before condensation 
breaks eventually down again. (a-c) Momentum distribution for states with 
condensates in three different modes. 
}
\label{fig:occups-hotbath}
\end{figure}

\emph{Excited-state condensation.}---%
Figure~\ref{fig:occups-hotbath}(d) shows how the occupations depend on the 
needle temperature $T_h$ for a system with $\gamma_h/\gamma=1$, $M=500$, 
$\ell=7$, and $T=2J$. Since the global temperature $T$ lies well above the 
equilibrium condensation temperature $T_c^\text{eq}\approx 0.05J$, no Bose
condensate is found at $T_h=T$, where the system is in equilibrium. 
However, when $T_h$ is increased, soon ground-state Bose condensation sets in.
When the needle temperature is increased further, remarkably a Bose condensate in 
the excited mode $k\approx \pi/\ell$ supersedes the ground-state condensate at a 
switch temperature $T_h^\text{s,1}$. The condensate mode switches once more to
$k\approx 2 \pi/\ell$ at $T_h^\text{s,2}$, before eventually at very large 
needle temperatures $T_h\gtrsim 10^4 J$ condensation breaks down again. The 
panels (a), (b), and (c) depict the momentum distribution for the three 
different needle temperatures marked in (d) and clearly show condensation in 
three different modes. 

We find that the switching of the condensate mode is triggered by the 
possibility to reduce the coupling between the condensate mode and the hot 
needle. Namely, typically the allowed wave numbers $k=\nu\pi/(M+1)$ that 
comply with the boundary conditions, do not assume those values $\kappa_\alpha$
that would perfectly decouple from the hot needle. We denote by $k_\alpha$ the
allowed wavenumber that minimizes the 
distance $\delta_\alpha = |k-\kappa_\alpha|$, which quantifies the coupling, 
$f^{(h)}_{qk_\alpha}\simeq 4\sin^2(\ell q)\ell^2\delta_\alpha^2$. 
With increasing $T_h$ we generally find a sequence of condensate modes
$k_0, k_{\alpha_1}, k_{\alpha_2} \ldots $, where $\alpha_{j+1}$ is the smallest 
value of $\alpha$ with $\alpha>\alpha_j$ and
$\delta_{\alpha}<\delta_{\alpha_j}$. The sequence ends, when the coupling 
cannot be lowered anymore by a larger $\alpha$. Since one has
$\delta_0=k_0$ for the ground state $k_0$ and fluctuating values
$\delta_\alpha \le \delta_0/2$ (depending on $M$ and $\ell$) for $\alpha\ge1$,
one always finds at least one switch of the condensate mode and $\alpha_1=1$. 
While for the parameters of Fig.~\ref{fig:mainfig}(b) the sequence ends 
already with $k_1$, it ends with $k_2$ for the parameters of
Fig.~\ref{fig:occups-hotbath} (for $\ell=21$ and $M=200$, we observe
the sequence $\alpha = 0,1,2,7$, not shown). 

We can estimate the switch temperature
$T_h^\text{s$\alpha$}$, above which a condensate in mode $k_\alpha$ forms. 
The momentum distribution can formally be written like 
$\la\no_k\ra = \sum_q R_{kq}\la\no_q\ra/B_k$ with 
$B_k=\sum_q[A_{qk}\la\no_q\ra+R_{qk}] $.
Physical occupations $\la\no_k\ra\geq0$ require $B_k>0$. 
Assuming a condensate with $\la\no_{k_\alpha}\ra\approx N$ particles in mode
$k_\alpha$, the temperature $T_h$ above which $\la\no_{k_{\alpha'}}\ra \geq 0$ 
for all $\alpha'<\alpha$ provides a good estimate for the switch temperature,
$T_{h}^\text{s$\alpha$}\approx 0.5n(\gamma/\ell\gamma_h)^2 
\max_{\alpha'<\alpha}(\Delta_{k_\alpha k_{\alpha'}}/\delta_{\alpha'}^2)$ 
\cite{Supp}, see dashed lines in Fig.~\ref{fig:occups-hotbath}(d) and
Fig.~\ref{fig:mainfig}(b). 
The linear dependence of $T_{h}^\text{s$\alpha$}$ on
$n$ indicates that excited-state condensation is suppressed (shifted to
$T_h = \infty$) in the limit of high densities, where ground-state condensation 
is expected for a system coupled to two thermal baths (of positive temperature) 
\cite{VorbergEtAl13, VorbergEtAl15}.
While we have presented examples for $\ell\ll M$, with the needle placed near 
the edge of the sample, and assumed this limit also in our analytical estimates, 
simulations show that the enhancement of the condensation temperature 
equally occurs for any needle position. However, placing the needle away from the 
edge we typically observe excited-state condensation, as it is indicated also by 
the $\ell^{-2}$ behavior of the estimated shift 
temperature $T_{h}^\text{s$\alpha$}$. 

\emph{General picture and conclusion}---%
An intuitive interpretation of the high-temperature and excited-state
Bose condensation observed here is that the nonequilibrium condensation can also
be viewed as a mechanism that suppresses the heat influx $I_h$ from the hot bath, 
with $I_b = \sum_{qk} \Delta_{qk} R_{qk}^{(b)}(\la\no_q\ra+1)\la\no_k\ra$.
Namely, keeping the distribution $\la\no_k\ra$ fixed, $I_h$ would increase linearly 
with $T_h$ [according to Eq.~(\ref{eq:rates})], while it still has to be balanced by
the outflux $I_g$ into the colder global bath (since $I_h=-I_g$ for a steady state).
This increase is prevented by forming of a condensate in a mode that almost 
decouples from the hot needle.
This interpretation is supported by Fig.~\ref{fig:occups}(e) showing that the
heat current $I=I_h=-I_g$ through the system plotted versus $T$ shows a maximum 
near the condensation temperature and, thus, a negative differential heat 
conductivity in the condensed regime. This counterintuitive effect is explained
by noting that the number of particles contributing to the heat transport is 
reduced by condensing into the ground state $k_0$, which hardly couples to the hot 
needle. The onset first of ground-state condensation and later also of
excited-state condensation observed when the needle temperature $T_h$ is increased 
[Fig.~\ref{fig:occups-hotbath}(d)] can, therefore, be understood as a strategy of 
the system to minimize its coupling to the hot needle (and with that $I_h$) further
and further. Note that these intriguing effects do not rely on the discrete nature 
of the tight-binding chain considered here and, therefore, occur equally in a 
continuous system \cite{Supp}.

The tendency to avoid a state with strong coupling to the environment bears 
resemblance to the quantum Zeno effect \cite{MisraSudarshan77}. The underlying 
mechanism, that the coupling to a second very hot bath can lead to a depletion of 
some system modes and, thus, to an enhanced occupation (and quantum degeneracy) of 
the subsystem defined by all other modes (not necessarily including the ground 
state), should be rather universal. It suggests a general strategy for the robust
preparation of quantum degenerate nonequilibrium states with unconventional 
properties and also in large-temperature environments by realizing
a hot bath coupling to all modes but a few. 
In future work, it will be 
interesting to explore such possibilities in bosonic and fermionic quantum 
systems. This includes the investigation of interacting systems and the role of 
stronger system-bath coupling.

\begin{acknowledgments}
\emph{Acknowledgment.}---%
We acknowledge discussion with Toni Ehmcke. This work was supported by the German 
Research Foundation DFG (FOR 2414). D.\ V.\ is grateful for support by the 
Studienstiftung des deutschen Volkes.
\end{acknowledgments}

\bibliography{mybib}

\end{document}


\title{
Supplemental material for ``High-temperature nonequilibrium Bose condensation induced by a hot needle"}
\author{Alexander~Schnell} 
\email[Electronic address: ]{schnell@pks.mpg.de}
\affiliation{Max-Planck-Institut f{\"u}r Physik komplexer Systeme, N{\"o}thnitzer Stra\ss e 38, 01187 Dresden, Germany}
\author{Daniel~Vorberg} 
\affiliation{Max-Planck-Institut f{\"u}r Physik komplexer Systeme, N{\"o}thnitzer Stra\ss e 38, 01187 Dresden, Germany}
\author{Roland~Ketzmerick} 
\affiliation{Max-Planck-Institut f{\"u}r Physik komplexer Systeme, N{\"o}thnitzer Stra\ss e 38, 01187 Dresden, Germany}
\affiliation{Technische Universit{\"a}t Dresden, Institut f{\"u}r Theoretische Physik and Center for Dynamics, 01062 Dresden, Germany}
\author{Andr{\'e}~Eckardt} 
\email[Electronic address: ]{eckardt@pks.mpg.de}
\affiliation{Max-Planck-Institut f{\"u}r Physik komplexer Systeme, N{\"o}thnitzer Stra\ss e 38, 01187 Dresden, Germany}

\date{\today}
\maketitle

In this supplemental material we (i) estimate the condensation temperature for a
one-dimensional tight-binding chain of finite length $M$ in equilibrium; 
(ii) justify the kinetic equations of motion obtained from the meanfield 
approximation by comparing it to Monte-Carlo results for the exact master 
equation; (iii) estimate the nonequilibrium condensation temperatures for the 1D and the 3D-like transition;
(iv) estimate the switch temperatures for excited state condensation;
and (v) show that the nonequilibrium  condensation phenomena found 
for a discrete tight-binding chain can also be observed in a continuous
one-dimensional system. 


\section{Condensation temperature in equilibrium}
\begin{figure}[t]
	\includegraphics[scale=1.2]{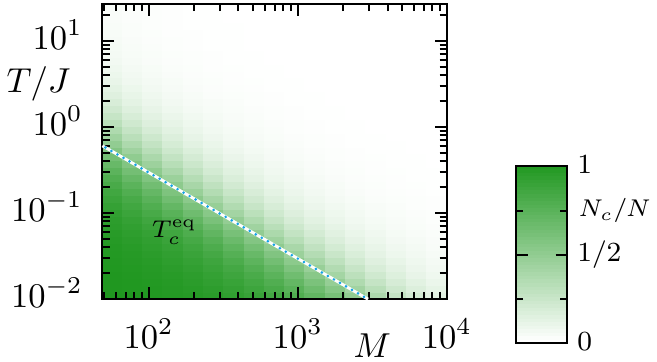}
	\caption{Condensate fraction $N_c/N$ for a tight binding chain of $M$ sites 
    at temperature $T$ (shading). The blue-white dotted line gives the 
    analytical estimate for the condensation temperature, where half of the 
    particles occupy the single-particle ground state.}
	\label{fig:equilibrium}    
\end{figure}

Under equilibrium conditions, when the system is coupled only to the global bath 
of temperature $T$ (i.e.\ for $\gamma_h=0$), Equation \refkineq\ (of the main text) is 
solved by the grand-canonical mean occupations
\be
\la\no_k\ra = \frac{1}{e^{ (\varepsilon_k-\mu)/T}-1}
\ee
with chemical potential $\mu$. 
When (finite-size) Bose condensation sets in, $\mu$ approaches $\varepsilon_{k_0}$ 
from below, so that the occupations of the low-energy modes with $k\ll1$ can be 
approximated by
\be
\la\no_k\ra \simeq  \frac{T}{\varepsilon_k-\mu} 
            \simeq \frac{T}{J k^2 -2J -\mu},
\ee
where we have used $\varepsilon_k =-2J\cos(k)\simeq -2J+Jk^2$. The chemical 
potential can be expressed in terms of the occupation $N_c=\la \no_{k_0}\ra$ of 
the ground state with wave number $k_0=\pi/(M+1)$, 
\be
\mu= -2J + Jk_0^2 -T/N_c.
\ee
For low temperatures, the number $N'$ of particles occupying excited states, 
with $k=\nu\pi/(M+1)$, is dominated by the long-wavelength modes, so that we can 
approximate
\be
N' = \sum_{k'\ne k_0} \la\no_k\ra
   \simeq \sum_{\nu=2}^\infty 
        \frac{1}{\frac{J\pi^2}{TM^2}(\nu^2-1)+\frac{1}{N_c}}.
\ee
For a finite system, we define the characteristic temperature $T_c$, where Bose 
condensation sets in, as the temperature for which half of the particles occupy 
the single-particle ground state, $N'=N_c=N/2$. It is given
by 
\be\label{eq:T_ceq}
T^\text{eq}_c \simeq \frac{a\pi^2}{2} \frac{nJ}{M} \approx  8.3\,\frac{nJ}{M},
\ee
where $a\approx1.68$ solves $1=a\sum_{\nu=2}^\infty 1/(\nu^2 + a-1)$. In
Fig.~\ref{fig:equilibrium} we plot the ground-state occupation (i.e.\
the condensate fraction) of the tight binding chain together with the estimate
(\ref{eq:T_ceq}) for the condensation temperature $T^\text{eq}_c$. The inverse 
dependence of $T^\text{eq}_c$ on the system size $M$ reflects the well-known 
result that in one spatial dimension, in the thermodynamic limit Bose-Einstein 
condensation is suppressed by thermal long-wavelength fluctuations.

\begin{figure}
	\includegraphics[scale=1.2]{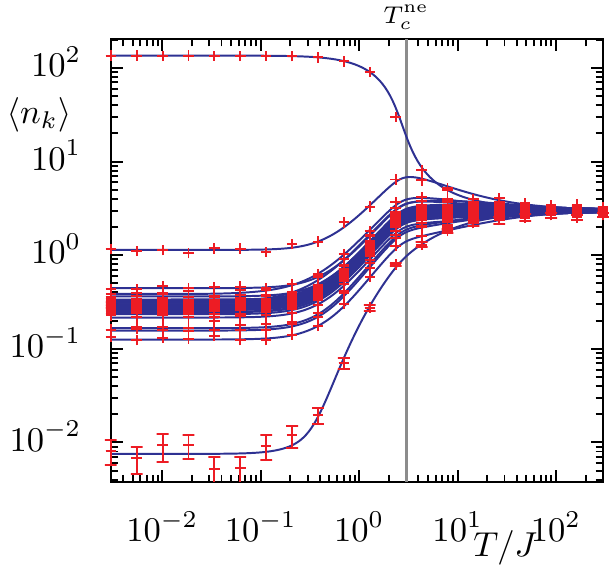}
	\caption{
    Mean occupations $\langle n_k \rangle$ for a system with $M=50$, $n=3$,
    $l=3$, $\gamma=\gamma_h$, $T_h = 120J$. Crosses represent the 
    Monte-Carlo results. We show error bars for the least occupied state only, 
    because for the other states they are too small to be visible. The data is 
    well approximated by the solid lines which result from the meanfield 
    approximation.}
	\label{fig:monte-carlo}
\end{figure}

\section{Quasiexact Monte-Carlo results for a small system}
The meanfield approximation $\la\no_q\no_k\ra\approx\la\no_q\ra\la\no_k\ra$, 
which gives rise to the closed set of kinetic equations~\refkineq\ for the 
mean occupations, allows us to treat large systems of up to $M=10^4$ lattice 
sites and to find analytical estimates for the parameters where condensation 
sets in. In order to justify this approximation, we have also simulated the 
full many-body rate equation for the probability distribution $p_\bn$ for 
finding the system in the eigenstate $|\bn\ra$. It reads
\be
\dot{p}_\bn = \sum_{kq} (1+n_q) n_k \big(R_{kq}p_{\bn_{q\leftarrow k}}-R_{qk}p_\bn\big),
\ee
where $\bn_{q\leftarrow k}$ denotes the vector of occupation numbers obtained from $\bn$ by 
transferring a particle from mode $k$ to mode $q$. 

An efficient way of solving this equation is given by quantum-jump Monte-Carlo 
simulations (see, e.g., \cite{PlenioKnight98}). For that purpose we generate a 
random walk in the classical space of Fock states $\bn$
(which is exponentially large with respect to the system size, but much smaller 
than the Fock space, which contains also the coherent superpositions of the 
Fock states). Namely, according to the sum and the relative weight of the
many-body rates $R_{qk}(n_q +1)n_k$ leading away from the current state $\bn$, 
we draw both the time after which a quantum jump happens and the new state
$\bn_{q\leftarrow k}$, respectively. Expectation values like the mean occupations
$\la\no_k\ra$ are computed by averaging over a random path. This method gives 
quasiexact results, in the sense that the accuracy is controlled by the length 
of the random path. A detailed description of the method is given in
Ref.~\cite{VorbergEtAl15}.

In Fig.~\ref{fig:monte-carlo} we plot the mean occupations $\la\no_k\ra$ of a 
system of $M=50$ sites and $n=3$, $\ell=3$, $\gamma_h/\gamma=1$, and
$T_h=120J$. The Monte-Carlo data (red crosses) are reproduced almost perfectly 
by the meanfield solution (solid lines). Already in this rather small system, 
we can see a rather sharp crossover, to a Bose condensed regime at
$T_c^\text{ne}$. 

\section{Nonequilibrium condensation temperature}
As described in the main text, in our system a Bose condensate in the ground state $k_0=\pi/(M+1)$ can be destroyed either by a large 
occupation of long-wavelength modes or by a large occupations of excited states at all energies. 
While the former case resembles condensation in a 1D system in equilibrium and leads to a system-size dependent 
condensation temperature $T_c^\text{ne1}$, the latter case bears similarity to equilibrium condensation in 3D and happens 
at a condensation temperature $T_c^\text{ne2}$, which is independent of the system size. 
Below, we will estimate both temperatures. The generalization to excited-state condensation is straightforward and not presented here.

\subsection{Condensation destroyed by long-wavelength modes}

Let us suppose that the system is in the condensed regime, where a 
large fraction $N_c/N$ of the particles occupy the ground-state mode
and the modes at large quasimomenta are approximately unoccupied (which requires $T\ll T_c^{\text{ne}2}$, see next section).
The mean occupations $ \la \no_k \ra$ of the long-wavelength modes $k$ with $k_0<k\ll 1$ in the vicinity of $k_0$ 
are then 
obtained from
\be
\label{eq:rate-1D-distri}
\begin{split}
	0 &\approx A^{(g)}_{kk_0}   \la \no_k \ra  N_c + R^{(g)}_{kk_0} N_c +\\
	 &\sum_{q\neq k_0} A_{kq}   \la \no_k \ra \la \no_q \ra+
	     \sum_{q\neq k_0} R_{kq}  \la \no_q \ra - \sum_q R_{qk} \la \no_k \ra,
\end{split}
\ee
which was derived from Eq.~\refkineq\ by neglecting the weak residual coupling of the ground-state mode to the hot needle, $R^{(h)}_{kk_0} \approx 0 \approx A^{(h)}_{kk_0}$ . 
The last term describes the coupling to approximately empty modes at larger quasimomenta $q$.

We can now use the assumption of a large needle temperature $T_h\gg T$. 
Employing the scaling $\la \no_k \ra \propto T/T_h$ (which will be shown to be self-consistent below), we find
\be
\begin{split}
	0 =& R^{(g)}_{kk_0} N_c +  \sum_{q\neq k_0} R^{(h)}_{kq}  \la \no_q \ra - \sum_q R^{(h)}_{qk}  \la \no_k \ra \\&+ \mathcal{O}(T/T_h),
\end{split}
\ee
where we obtained the scaling of $R_{kq}$ and $A_{kq}$ from Eqs.~\refeqrates\ and \refeqA\ from the main text.
The rates $R^{(h)}_{kq}$  scale like $k^2$ for small $k$. So for $k$ going to zero, we may omit the second term, but not the third one since
$\la \no_k \ra$ is peaked around zero, which will keep the third term finite. This yields
\bes
0 \approx & R^{(g)}_{kk_0} N_c -\sum_q R^{(h)}_{qk} \la \no_k \ra \\
    \approx & \,
   \twopihbar  \gamma^2 T N_c- \twopihbar 2\gamma_h^2\ell^2 k^2 M T_h \, \la n_k \ra .
\ees
For the second approximation we 
used the small-$|\Delta_{kq}|/T$ expression, Eq.~\refeqrates, for the rates $R^{(g)}_{kk_0}$ and $\sum_q R^{(h)}_{qk}\approx\twopihbar 4 \gamma_h^2 T_h \sin^2(\ell k)
\sum_q\sin^2(\ell q)\approx \twopihbar 2\gamma_h^2\ell^2 k^2 M T_h$. Thus, for large $T_h$ we have
\be
	\label{eq:nklw}
   \la \no_k \ra = \frac{\twopihbar N_c \gamma^2 T}{\twopihbar M 2\gamma_h^2\ell^2 T_h k^2} = \frac{N_c w^2}{2 M} \frac{1}{k^2},
\ee
with $w$ as defined in the main text. The occupations scale like $T/T_h$  as required by self consistence. 
Even though expression \eqref{eq:nklw} is derived for modes $k$ with $k_0<k\ll1$, we can use it for all $k\ne k_0$ since it vanishes rapidly with increasing~$k$.
With that, the condensate
depletion due to the long wave modes $N'=\sum_{k\ne k_0} \la \no_k \ra $ is approximately given by
\begin{equation}
	N' \approx N_c w^2 \frac{M}{2\pi^2}\sum_{\nu=2}^\infty \frac{1}{\nu^2}.
\end{equation}
Using the definition that at the condensation temperature $N_c=N'=N/2$, we can estimate
\begin{equation}
T_c^\text{ne1}\approx 30.6 \frac{\ell^2\gamma_h^2T_h}{\gamma^2} \,\frac{1}{M}.
\end{equation}

\subsection{Condensation destroyed by modes with large quasimomentum}
Let us again suppose that the system is in the condensed regime with a large fraction $N_c/N$ 
occupying the ground-state mode. But now, we assume that $T\ll T_c^{\text{ne}1}$ so that the long-wavelengths modes are hardly occupied. 
Furthermore, we assume that the occupations of the most of the excited modes $k\ne k_0$ 
are close to a flat distribution corresponding to the hot temperature $T_h\gg J$, so that we can approximate $\la\no_k\ra\approx N'/M$ for $k \neq k_0$. 
Plugging this ansatz into the rate equation \refkineq\ we find for $k=k_0$
\be
\begin{split}
	0 \approx  \frac{N'}{M}  N_c \sum_{q} A^{(g)}_{k_0q} + \frac{N'}{M} \sum_{q\neq k_0} R^{(g)}_{k_0q}  - N_c \sum_{q\neq k_0} R^{(g)}_{qk_0}.
\end{split}
\ee
Analogous to Eq.~\eqref{eq:rate-1D-distri}, here we assumed that the ground state $k_0$ decouples from the needle.
In the condensed regime we find $N_c \gg T/J$, and $N_c\gg N'/M$ and therefore the first  and the last term dominate over the second one.
Using $\sum_{q} A^{(g)}_{k_0q} = \gamma^2 \sum_q\Delta_{qk_0}\approx \gamma^2 2J M$ this leads to
\be
 0 \approx  \gamma^2 \twopihbar N_c \, \biggl[2J N' - T  s(T)\biggr],
 \ee
where we have defined the sum 
\be
	\label{eq:sumT}
	s(T) =  \sum_{q\neq k_0} \frac{\Delta_{qk_0}/T}{e^{\Delta_{qk_0}/T}-1}\ \overset{T \gg J}{\longrightarrow}\ M-1.
\ee
So we find the total depletion to be given by
\be
	N' =  \frac{T  s(T)}{2J}.
\ee
Solving this equation for $N_c=N'=nM/2$, 
we may obtain $T_c^\text{ne2}$ numerically. 
However, with the limiting value of $s(T_c^\text{ne2}) \approx M-1$ from Eq.~\eqref{eq:sumT}
the condensation temperature reads $T_c^\text{ne2} \approx nJ$, which is consistent with the assumption $T_c^\text{ne2}\gg J$ as long as $n\gg1$. 

\section{Temperature for switch to excited state condensation}

Let us now discuss, under which conditions we can expect condensation in an excited state, $k_c = k_\alpha \approx \frac{\pi}{\ell}, \frac{2\pi}{\ell}, \dots$,  in the finite system.
The mean field equation \refkineq\
must hold for all $k$, which we may solve formally like
\begin{equation}
	 \langle n_k \rangle =\frac{\sum_q R_{kq} \langle n_q \rangle }{\sum_q \bigl( A_{qk} \langle n_q \rangle + R_{qk}  \bigr)}.
	 \label{eq:nk-explicit}
\end{equation}
We will use the fact that all occupations must be non-negative, $\langle n_k \rangle \geq 0$, to
find the regime where excited state condensation may occur.

Since the numerator of Eq.~\eqref{eq:nk-explicit} is strictly positive, the denominator has to
be positive, too. Deep in the condensed regime, we may neglect the depletion, $N_c\approx N$, which yields
\begin{equation}
	A_{k_c k} N  + \sum_{q} R_{qk} > 0.
\end{equation}
Considering that the condensate decouples from the hot bath, the rate imballance $A_{k_ck}$ is dominated
by the global bath and we can approximate $A_{k_c k} \approx  \gamma^2 \Delta_{kk_c}$.
Additionally $T_h \gg T$ holds, such that $\sum_{q} R_{qk} \approx \sum_{q} R^{(h)}_{qk} \approx \gamma_h^2 T_h 2M \sin(k \ell)^2$.
In total, we have to satisfy
\begin{equation}
	\gamma^2 \, \Delta_{kk_c} {N} + \gamma_h^2 T_h 2{M} \sin(k \ell)^2  > 0
	\label{eq:positivity-condition}
\end{equation}
for all $k$.
For ground-state condensation $k_c = k_0 ={\pi}/({M+1})$ both terms are positive, such that for all parameter values the ground state condensate ansatz
does not violate the physical requirement of positive occupations. For $k_\alpha \approx \kappa_\alpha= \frac{\alpha \pi}{\ell}$ with $\alpha \geq 1$ however, the first term is negative
 for all $k < k_\alpha$ and must be compensated by the second term. Since the
second term is minimal at one of the decoupling wavenumbers $k_{\alpha'}$, $\alpha' < \alpha$, it suffices to check for positivity at these wave numbers. From this requirement, we find that for
\begin{equation}
	T_h > T^{s,\alpha}_h =\frac{n\gamma^2}{2(\ell\gamma_h)^2} \ \max_{\alpha'<\alpha} \left(\frac{\Delta_{k_\alpha k_{\alpha'}}}{\delta_{\alpha'}^2}\right)	
	\label{eq:temperature-scale}
\end{equation} 
condensation in the excited state $k_\alpha$ may occur.

\section{Continuous system}
\begin{figure}[t]
\includegraphics[scale=1.2]{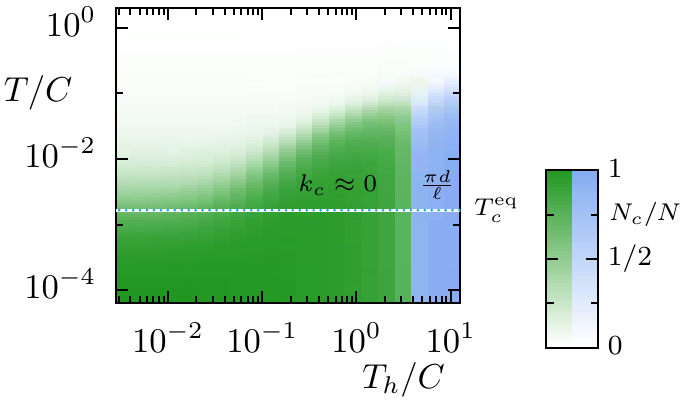}
\caption{Condensate fraction $N_c/N$ for the continuous 1D system of length
$L=500d$ and density of $n=0.1/d$ versus the temperatures $T$ and $T_h$ of the 
global bath and the hot needle, respectively. The spatial extent of the hot 
needle is $d$ and it is placed at distance $\ell = 20 d$ from the edge. The 
relative coupling between both baths is $\gamma_h/\gamma =1/4$. Green (light 
blue) shading indicates the relative number of particles in the mode
$k_0\approx 0$ ($k_1\approx\pi d/\ell $). The blue-white dotted line 
corresponds to the condensation temperature in equilibrium.}
	\label{fig:box}
\end{figure}

In the main text, we treated a one-dimensional (1D) system of noninteracting bosons 
in a tight-binding lattice. However, the effects discussed in the main text do 
not depend on the discrete nature of the tight-binding lattice. They occur in a 
similar form also in a continuous 1D system described by the
Hamiltonian
\be
\hat{H} = \int_0^L \!\rd x \,\hat{\psi}^\dag(x)
        \bigg(-\frac{\hbar^2}{2m}\frac{\rd^2 }{\rd x^2}\bigg) \hat{\psi}(x)
        = \sum_k \varepsilon_k\no_k,
\ee
where $m$ denotes the mass of the particles and $\hat{\psi}(x)$ the field 
operator annihilating a boson at position $x$. The dimensionless wavenumbers
$k= \nu\pi d/L$ with $\nu=1,2,\ldots$ and some length scale $d$ (which we take
to be the extent of the hot needle defined below) characterize the single-particle 
eigenstates with energy $\varepsilon_k=\frac{\hbar^2 k^2}{2md^2}\equiv C k^2$, 
and wave functions $\la x|k\ra=\sqrt{2/L}\, \sin(kx/d)$. The corresponding number 
operator reads $\no_k=\ca_k\co_k$ with
$\co_k = \int_0^L \!\rd x\, \la k|x\ra \, \hat{\psi(x)}$. The energy eigenstates 
of the system are Fock states $|\bn\ra$ labeled by the vector of occupation 
numbers $n_k$. 

In such a continuous system, a hot local bath at position $x=\ell$ of spatial 
extent $d$, can be described by the coupling operator
\be
\hat{v}_h=\frac{L}{d}\int_{\ell-\frac{d}{2}}^{\ell+\frac{d}{2}} \!\rd x \,
                        \hat{\psi}^\dag(x)\hat{\psi}(x). 
\ee
The corresponding single-particle rates are of the form of Eq.~\refeqrates, 
with $f^{(h)}_{kq}=[\cos(k\frac{\ell}{d})\,\mathrm{sinc}(k/2)-\cos(q\frac{\ell}{d})\, \mathrm{sinc}(q/2)]^2$ 
giving $f^{(h)}_{kq}\simeq 4\sin^2(k{\ell}/{d})\sin^2(q{\ell}/{d})$ in the limit $d\to0$. A global 
heat bath is still described by Eq.~\refeqrates with $f^{(g)}_{kq}=1$.

From these rates we computed the corresponding mean-occupations for a system of 
$N=0.1 L/d$ particles, length $L=500 d$, and needle position $\ell = 20d$. In
Fig.~\ref{fig:box} we plot the condensate fraction, i.e. the fraction of 
particles occupying the most occupied mode whose wave number is indicated by 
the color of the shading (green for $k_0\approx 0$ and light blue for
$k_1\approx \pi d/\ell$). As for the tight-binding chain, we can clearly see 
that the temperature of the global bath at which the system condenses increases 
with needle temperature. As a result Bose condensation is found for a system 
coupled to two baths both having temperatures well above the equilibrium 
condensation temperature $T_c^\text{ne}$ (which is indicated as blue-white dotted 
line). Moreover, when the needle temperature is increased 
further, ground-state condensation in mode $k_0\approx 0$ is superseded by the 
formation of a condensate in the excited mode $k_1\approx \pi d/\ell$, which 
provides a better decoupling from the hot needle.

\bibliography{mybib}